\documentclass[review,3p]{elsarticle}
\usepackage{graphicx}
\usepackage{amssymb}
\usepackage{lineno}
\usepackage{color}
\usepackage{verbatim}

\begin{document}

\title{Transverse spin-dependent azimuthal correlations of charged pion pairs measured in p$^\uparrow$+p collisions at $\sqrt{s}$ = 500 GeV}

\author{
L.~Adamczyk$^{1}$,
J.~R.~Adams$^{29}$,
J.~K.~Adkins$^{19}$,
G.~Agakishiev$^{17}$,
M.~M.~Aggarwal$^{31}$,
Z.~Ahammed$^{54}$,
N.~N.~Ajitanand$^{42}$,
I.~Alekseev$^{15,26}$,
D.~M.~Anderson$^{44}$,
R.~Aoyama$^{48}$
% ...
}

\address{$^{1}$AGH University of Science and Technology, FPACS, Cracow 30-059, Poland}
% ...
\address{$^{15}$Alikhanov Institute for Theoretical and Experimental Physics, Moscow 117218, Russia}
% ...
\address{$^{17}$Joint Institute for Nuclear Research, Dubna, 141 980, Russia}
% ...
\address{$^{19}$University of Kentucky, Lexington, Kentucky 40506-0055}
% ...
\address{$^{26}$National Research Nuclear University MEPhI, Moscow 115409, Russia}
% ...
\address{$^{29}$Ohio State University, Columbus, Ohio 43210}
% ...
\address{$^{31}$Panjab University, Chandigarh 160014, India}
% ...
\address{$^{42}$State University of New York, Stony Brook, New York 11794}
% ...
\address{$^{44}$Texas A\&M University, College Station, Texas 77843}
% ...
\address{$^{48}$University of Tsukuba, Tsukuba, Ibaraki, Japan,305-8571}
% ...
\address{$^{54}$Variable Energy Cyclotron Centre, Kolkata 700064, India}
% ...

\date{\today}

\begin{abstract}
The transversity distribution, which describes transversely polarized quarks in transversely polarized nucleons, is a fundamental component of the spin structure of the nucleon, and is only loosely constrained by global fits to existing semi-inclusive deep inelastic scattering (SIDIS) data. In transversely polarized $p^\uparrow+p$ collisions it can be accessed using transverse polarization dependent fragmentation functions which give rise to azimuthal correlations between the polarization of the struck parton and the final state scalar mesons.

This letter reports on spin dependent di-hadron correlations measured by the STAR experiment. The new dataset corresponds to 25 pb$^{-1}$ integrated luminosity of $p^\uparrow+p$ collisions at $\sqrt{s}=500$~GeV, an increase of more than a factor of ten compared to our previous measurement at $\sqrt{s}=200$~GeV.  Non-zero asymmetries sensitive to transversity are observed at a $Q^2$ of several hundred GeV and are found to be consistent with the former measurement and a model calculation. %we observe consistent with the former measurement are observed.} 
We expect that these data will enable an extraction of transversity with comparable precision to current SIDIS datasets but at much higher momentum transfers where subleading effects are suppressed.
\end{abstract}


\maketitle

\section{Introduction}
The proton is the fundamental bound state of quantum chromodynamics (QCD). In spite of its importance for our understanding of this theory, our knowledge of the proton structure remains incomplete~\cite{Geesaman:2015fha}. In particular, the proton wave function cannot be computed ab-initio in perturbative QCD (pQCD), but has to be constrained by measurements.
In deep inelastic scattering (DIS) experiments of electrons or muons off nuclei at high energies, the wavefunction of the proton is accessed on the lightcone. In this frame, the wavefunction can be expanded in the squared 4-momentum transfer $Q^2$ of the interaction. The leading coefficients in this expansion can be identified with three parton distribution functions (PDFs). In the parton model, PDFs have a probabilistic interpretation as the probability of finding a parton that carries a momentum fraction $x$ of the parent proton. 
%The PDFs are functions of $x$, which is also known as the Bj\"orken scaling variable and $Q^2$.  
The moderate $Q^{2}$ dependence, which arises from the parton splitting functions~\cite{Altarelli:1977zs,Gribov:1972,Dokshitzer:1977}, is computed using evolution equations.  We assume a $Q^2$ dependence in the following discussion even when not explicitly written.
Two of the PDFs, the parton helicity averaged PDF $f_1(x)$, and the helicity PDF $g_1(x)$ appear at leading twist respectively in the spin averaged and longitudinally polarized inclusive DIS cross-section \cite{Mulders:1995dh}. They are therefore  fairly well determined experimentally~\cite{Aidala:2012mv}. The third one, the transversity distribution $h_1(x)$, does not appear at leading twist in the inclusive DIS cross-section since it is connected to a chiral-odd helicity-flip amplitude. Instead, it is accessed in processes where it couples to the chiral-odd transverse spin dependent fragmentation function (FF)~\cite{Metz:2016swz}. 
%At leading order in the negative squared 4-momentum transfer $Q^2$ and neglecting intrinsic transverse momenta, the nucleon spin structure probed in deep-inelastic-scattering (DIS) can be described by three so-called parton distribution functions (PDFs) which are functions of the Bjorken scaling variable $x$. Two of these, the spin integrated PDF $f_1(x)$ and the helicity PDF $g_1(x)$ can be probed in inclusive DIS and are therefore  fairly well known~\cite{Aidala:2012mv}. The third one, the so-called transversity distribution $h_1(x)$ does not appear at leading order in $Q^2$ in the DIS cross-section since it is connected to a chiral-odd helicity-flip amplitude. Instead, it can only be accessed in a process where it couples to another chiral-odd quantity, which for all practical purposes is a transverse spin dependent fragmentation function (FF)~\cite{Metz:2016swz}. 
The transversity PDF can be interpreted as the probability of finding a transversely polarized quark in a transversely polarized proton, and the FF serves as a quark polarimeter. 

\begin{figure} %[h]
\centering
\includegraphics[width=12cm]{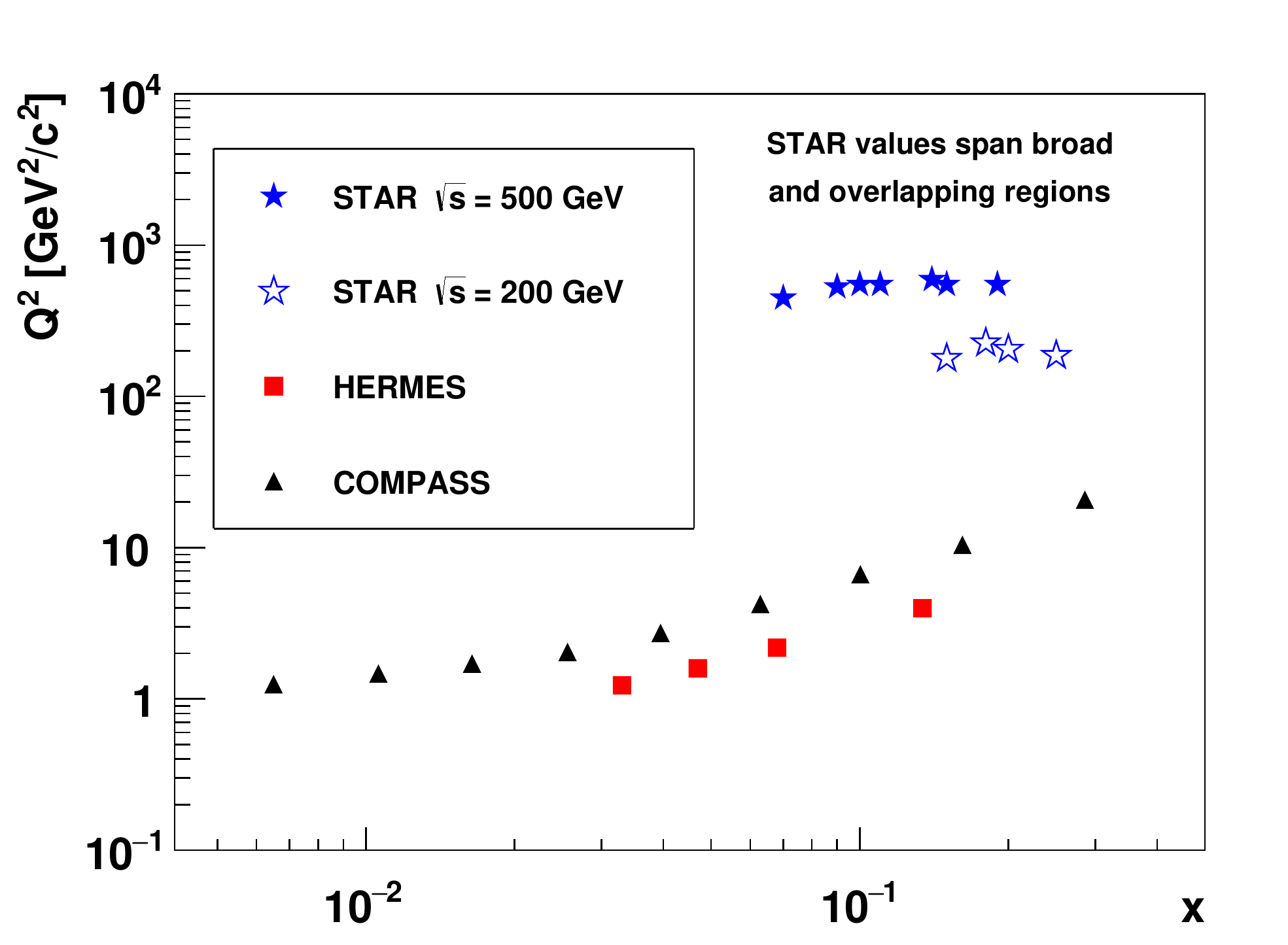}
\caption{$Q^{2}$ vs $x$ coverage for STAR, HERMES, and COMPASS~\cite{Airapetian:2004tw,Airapetian:2008sk,Adolph:2012nw,Alekseev:2010rw,Alekseev:2008aa}. The kinematics of the STAR data points correspond to the lower panel of Fig. \ref{f2}.\label{Q2vsX}}
\end{figure}

The analysis presented here investigates a channel in which transversity couples to the spin dependent di-hadron FF $H_1^\sphericalangle(z,M)$~\cite{Bianconi:1999uc,Jaffe:1997hf,Boer:2003ya}, which, for historical reasons, is also known as the interference fragmentation function (IFF). Here, $z$ is the fraction of the parent parton energy carried by the hadron pair, and $M$ is the invariant mass of the pair.
Presently, transversity is only loosely constrained by fits~\cite{Kang:2015msa, Radici:2015mwa} to available SIDIS~\cite{Airapetian:2004tw,Airapetian:2008sk,Adolph:2012nw,Alekseev:2010rw,Alekseev:2008aa} and $e^+e^-$~\cite{Abe:2005zx,Vossen:2011fk} data. The $e^+e^-$ data are necessary to constrain the polarization dependent fragmentation functions. While measurements sensitive to the unpolarized single hadron fragmentation functions have a long history (see again~\cite{Metz:2016swz} for an overview), only recently, a result sensitive to the unpolarized di-hadron fragmentation function~\cite{Seidl:2017qhp} was presented for the first time.  
Fixed target data are currently limited in the valence region to $x<0.2$, restricting the knowledge of valence quark transversity at high $x$. Probing transversity in $p+p$ collisions provides better access to the $d$-quark transversity than is possible in SIDIS, due to the fact that there is no charge weighting in the hard scattering QCD $2\rightarrow 2$ processes in p+p collisions.
%Due to the lack of precision data on effective neutron targets, our knowledge of the $d$ quark transversity is particularly poor.  
A precision determination of both $u$ and $d$-quark transversity are important in particular for the determination of the zeroth moment of transversity, the tensor charge
\begin{equation}
g_T=\int_{0}^1 dx[h_1^q(x)-h_1^{\bar{q}}(x)]
\end{equation}
Recently, $g_T$ has attracted increased interest. One reason is that it can be calculated precisely using lattice QCD~\cite{Bhattacharya:2016zcn,Green:2012ej,Abdel-Rehim:2015owa,Aoki:2010xg,Bali:2014nma}, which makes it one of the few observables involving transverse polarization where experiments can be compared with first principles pQCD calculations. In fact, $g_T$ is the first nucleon matrix element that could be extrapolated to the physical limit. Furthermore, $g_T$ determines the effective tensor coupling constant for beyond the standard model contributions to low energy scattering~\cite{Bhattacharya:2011qm}. This determination is particularly important for planned electric dipole moment experiments where a precise knowledge of $g_T$ is needed to determine the contributions of possible new CP violating phases~\cite{Dubbers:2011ns}.
Due to its chiral-odd property, gluon polarization contributions to transversity in a spin-$\frac{1}{2}$ target vanish~\cite{Barone:2001sp}. This characteristic is one reason $g_T$ is dominated by the medium to high $x$ region. 
%Experimentally, the vanishing contribution is advantageous because the low-$x$ region, where gluon contributions would rise dramatically, does not have to be mapped out. 
%However, the data in the valence region $x>0.2$ is currently sparse.
Precision data from transversely polarized $p+p$ collisions at high $\sqrt{s}$ and $p_{T}$ are crucial to access transversity at high $Q^2$, where theoretical uncertainties are well under control.  The kinematic region covered by the STAR experiment at these energies overlaps the reach of current SIDIS experimental data on transversity in the upper part of the covered $x$ range (see Fig. \ref{Q2vsX}). The STAR kinematics is obtained from the transverse momentum of the mid-rapidity jet containing the hadron pair since this is the relevant scale in $p+p$ collision and approximately equal to $Q^2$.
%The approximate STAR $Q^2$ values are inferred point-to-point by $(p_T^{pair}/z)^{2}$.}  Since the underlying partonic process is dominated by quark-gluon scattering, the $u$- and  $d$-quark transversity distributions are accessed equally.
The results presented in this letter at $\sqrt{s}=500$~GeV use more than 10 times the integrated luminosity than our previously reported result at $\sqrt{s}=200$~GeV \cite{Adamczyk:2015hri}, where a significant signal of transversity was observed in an exploratory measurement of di-pion correlations.
%One important finding was the confirmation that universality holds where the phase space of the $pp$ and the SIDIS data overlapped~\cite{Radici:2016lam}. Since the calculations can be performed in a collinear framework, this was already postulated. 
The calculations reported in \cite{Radici:2016lam} found hints of universality where the phase space of the $\sqrt{s}=200$~GeV $p+p$ and the SIDIS data overlap.  Since the calculations are performed in a collinear framework, this was already postulated.
However, since factorization is not proven in this process and has been explicitly shown to be broken in other transverse polarization dependent processes in $p+p$~\cite{Rogers:2010dm}, this was a crucial finding to support the inclusion of the data in global analyses. In the future, a comparison between di-hadron asymmetries, with measurements of azimuthal asymmetries of pions in jets by STAR \cite{Adamczyk:2017wld}, will provide further tests of universality and factorization. The former asymmetries can be described in a collinear framework, while the latter include an explicit dependency on intrinsic transverse momenta (for more details see \cite{Kang:2017glf,Kang:2017btw}). The collinear framework is well understood and describes the unpolarized $p+p$ cross-section well~\cite{Adare:2015ozj}, but the transverse momentum dependent (TMD) framework is still being developed, and questions remain about universality, factorization and evolution.

\section{Experiment}
The Relativistic Heavy Ion Collider (RHIC), located at Brookhaven National Laboratory, can collide beams of polarized protons, as well as heavy ions, at each of the interaction regions. %collides beams of protons at interaction regions situated around the RHIC ring.  
The data used in this analysis were recorded at the STAR experiment in 2011 representing 25 pb$^{-1}$ integrated luminosity of transversely polarized $p+p$ collisions at $\sqrt{s}=500$~GeV and an average beam polarization of 53$\%$.  Kinematic observables of charged particles are measured using the Time Projection Chamber (TPC) with $2\pi$ azimuthal coverage in the pseudorapidity range -1 $\lesssim \eta \lesssim$ 1 \cite{Ackermann:2002ad}.  The barrel and endcap electromagnetic calorimeters (BEMC/EEMC) and the beam-beam counters (BBC) are used in coincidence for the trigger.  A single BEMC tower is required to have a minimum transverse energy ($E_{T}>$ 4.0 or 5.7 GeV) or a $\Delta\phi\times\Delta\eta$ = 1.0$\times$1.0 jet patch must have $E_{T} >$ 6.4, 9.0 or 13.9 GeV, respectively. Particles are identified by measuring their average specific ionization energy loss,  $\left\langle dE/dx \right\rangle$, as they traverse the TPC and comparing this measured value with the associated parameterized expectation for each particle species as a function of $\eta$ and momentum.  Cuts on the number of standard deviations from the pion $\left\langle dE/dx \right\rangle$ peak (-1$\sigma$ to $2\sigma$) and the number of hits used to determine $\left\langle dE/dx \right\rangle$ ($>$ 20) are applied to achieve an 85$\pm 2.5\%$ pion pair purity across the entire kinematic range.  The pion pair purity is the probability that both particles in a pair are pions.  The momentum, $p$, of each particle is required to be greater than 2~GeV/\textit{c}.  

Each proton beam in the RHIC ring consists of bunches that alternate between being transversely polarized up or down with respect to the accelerator plane.  However, when the single spin asymmetry measurement is carried out with respect to a given beam, the polarization of the other beam is integrated over to effectively be unpolarized.  Polarimeters, which measure the elastic scattering of protons on ultra thin carbon ribbon targets several times during a fill, were used to measure the polarization of each beam.  These polarimeters were calibrated with a polarized hydrogen gas jet target \cite{polar}.     

\begin{figure} %[t]
\centering
\includegraphics[width=10cm]{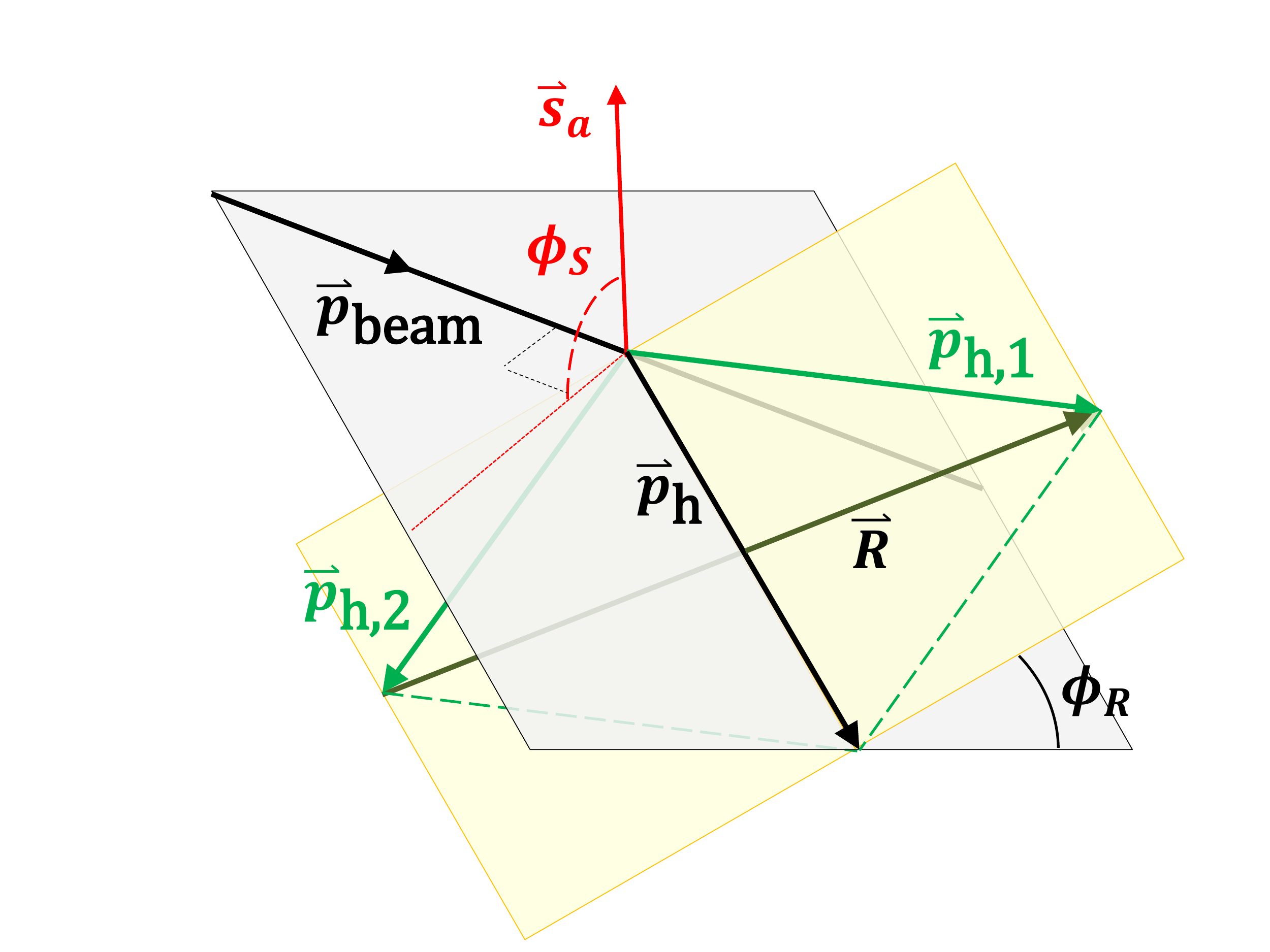}
\caption[]{Diagram of the azimuthal angle, where $\vec{p}_{h,1(2)}$ is the momentum of the positive (negative) pion, $\vec{s}_{a}$ is the beam polarization, and $\phi_{R}$ is the angle between the scattering plane (gray) and the di-hadron plane (yellow).\label{f1}}
\end{figure}

\begin{figure} %[pb]
\centering
\includegraphics[width=10cm]{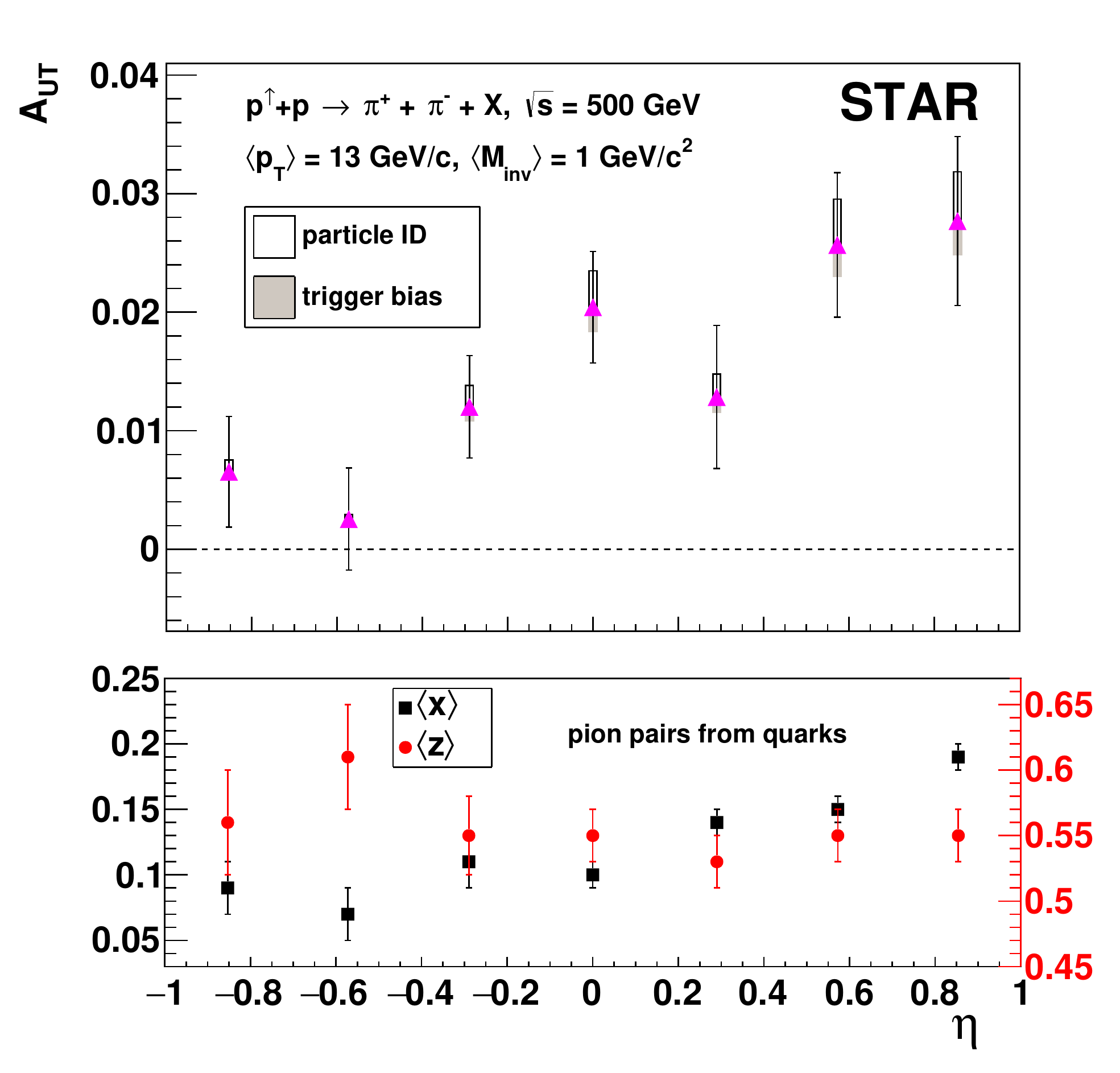}
\caption{$A_{UT}$ (top) and the kinematic variables, $\langle x\rangle$ and $\langle z\rangle$ (bottom), plotted as a function of $\eta$ for $\langle p_{T}\rangle$ = 13 GeV/\textit{c} for pairs that arise from quarks.  Statistical uncertainties are represented by the error bars, the open rectangles are the systematic uncertainties originating from the particle identification, and the solid rectangles represent the trigger bias systematic uncertainties.\label{f2}}
\end{figure}

\begin{figure*} %[t]
\includegraphics[width=18cm]{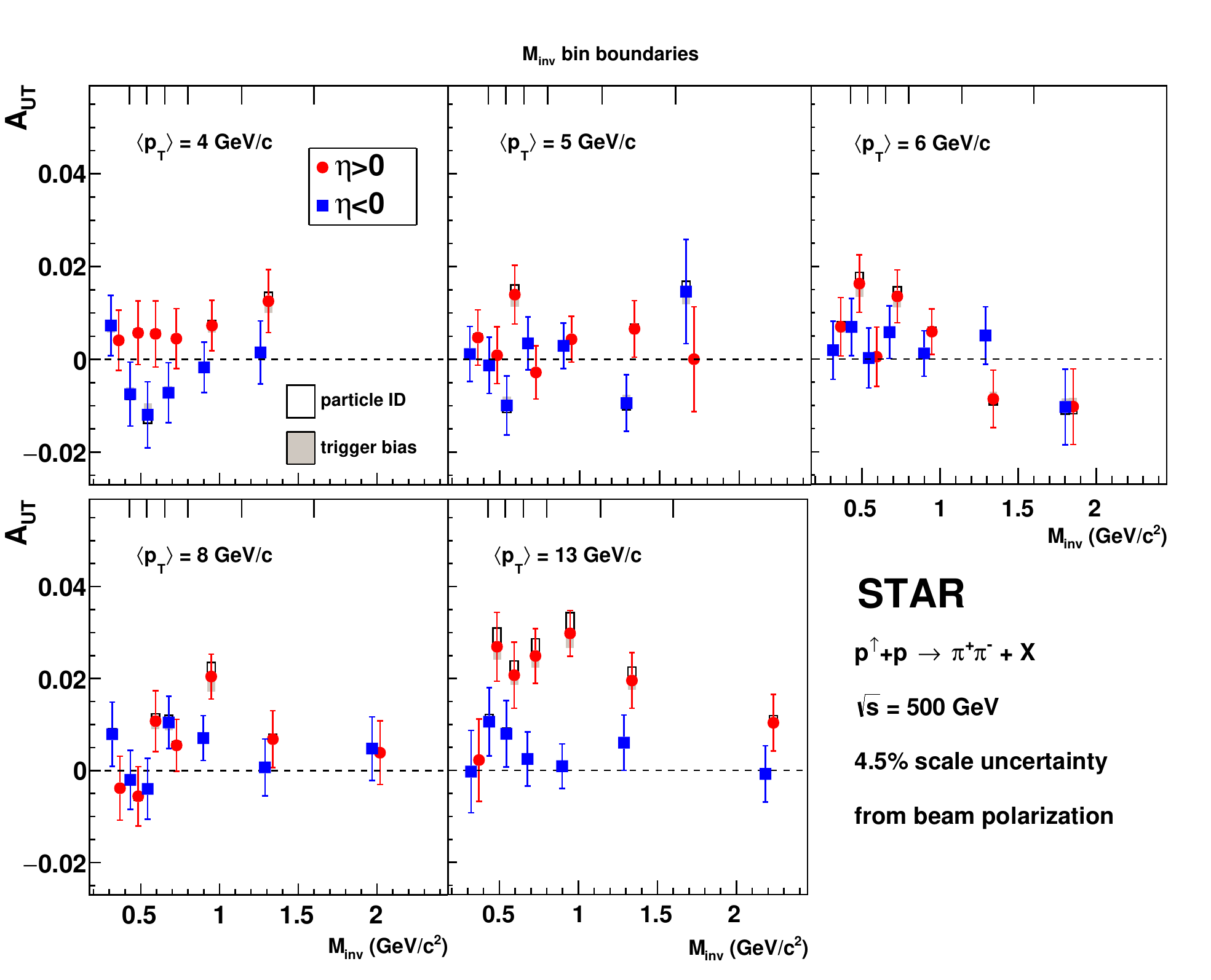}
\caption{The asymmetry $A_{UT}$ as a function of $M_{inv}$ for five $p_{T}$ bins. Statistical uncertainties are represented by the error bars, the open rectangles are the systematic uncertainties originating from the particle identification, and the solid one represent the trigger bias systematic uncertainties. 
The $M_{inv}$ bin boundaries are shown at the top of each panel.\label{f3}}
\end{figure*}

\begin{figure*} %[t]
\centering
\includegraphics[width=16cm]{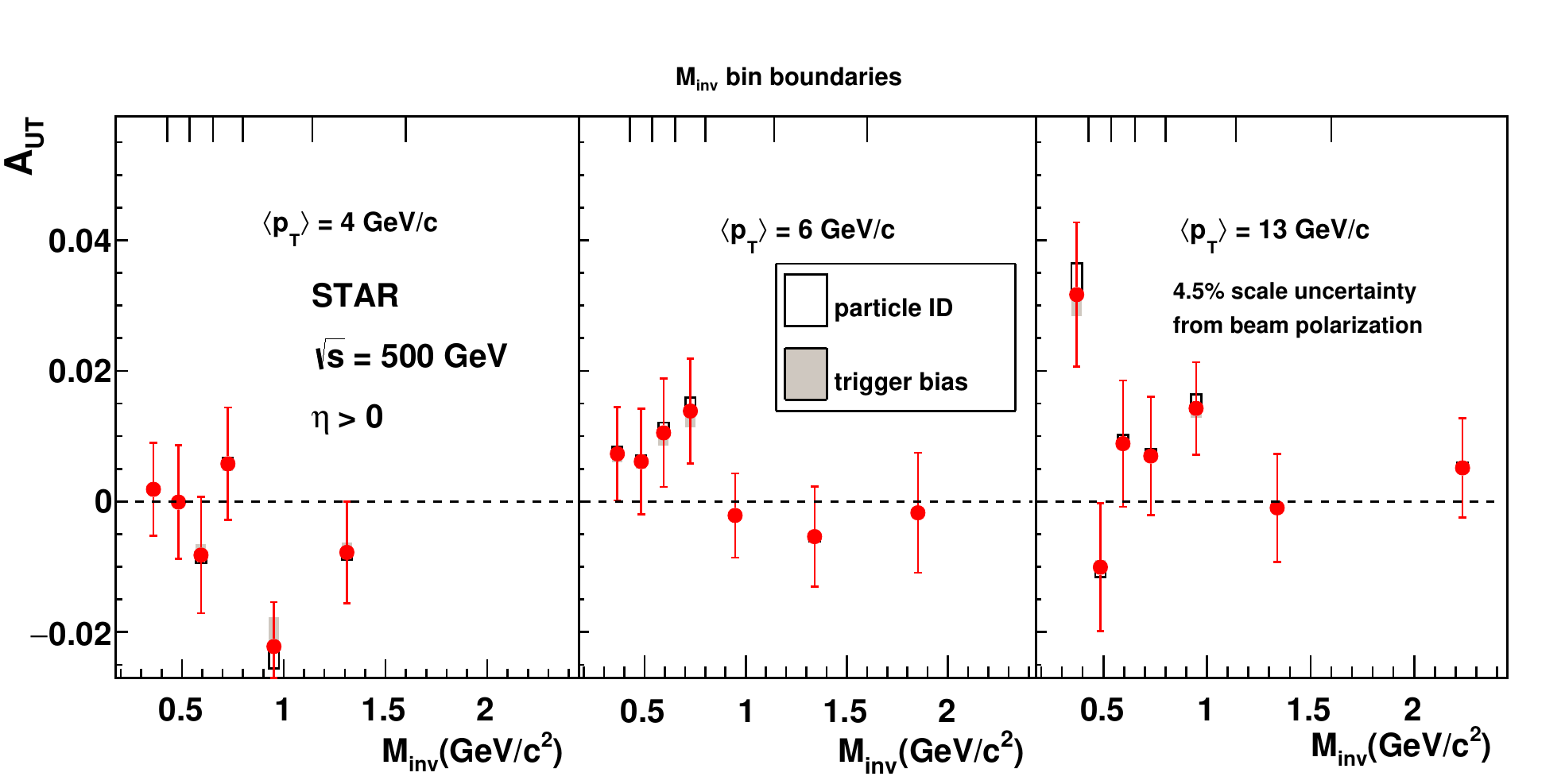}
\caption{The same-charge, momentum-ordered ($|\vec{p}_{h,1}| > |\vec{p}_{h,2}|$) asymmetry $A_{UT}$ as a function of $M_{inv}$ for the lowest $p_{T}$ bin, 
mid-$p_{T}$ bin, and the highest $p_{T}$ bin used in Fig.~\ref{f3}. Statistical uncertainties are represented by the error bars, the open rectangles are the systematic uncertainties originating from the particle identification, and the solid one represent the trigger bias systematic uncertainties. The $M_{inv}$ bin boundaries are shown at the top of the figure.\label{f5}}
\end{figure*}

\begin{figure*} %[b]
\includegraphics[width=18cm]{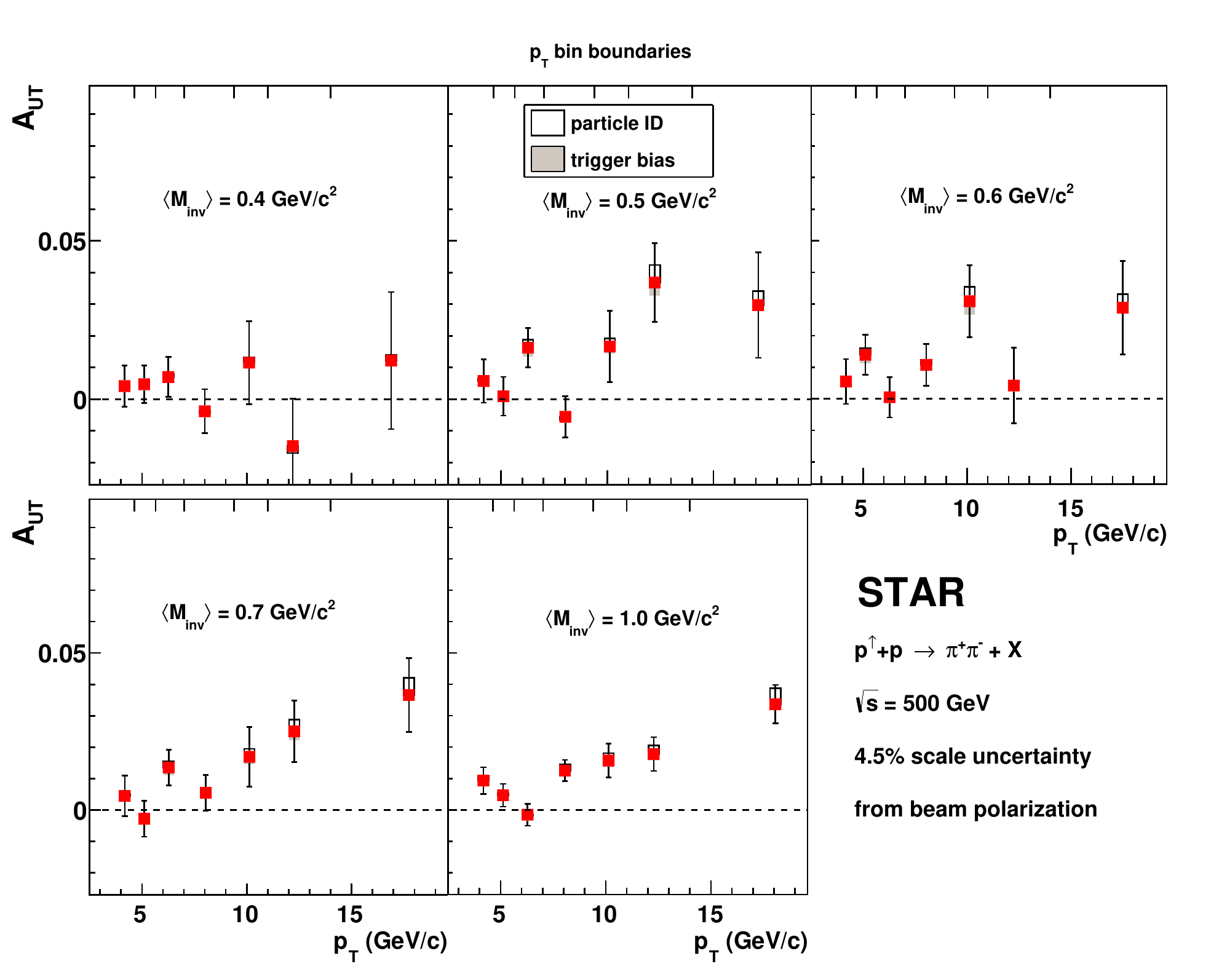}
\caption{The asymmetry $A_{UT}$ as a function of $p_{T}$ for five $M_{inv}$ bins for $\eta > 0$. Statistical uncertainties are represented by the error bars, the open rectangles are the systematic uncertainties originating from the particle identification, and the solid one represent the trigger bias systematic uncertainties. The $p_{T}$ bin boundaries are shown at the top of the figure.\label{f4}}
\end{figure*}

\begin{figure} %[h]
\centering
\includegraphics[width=12cm]{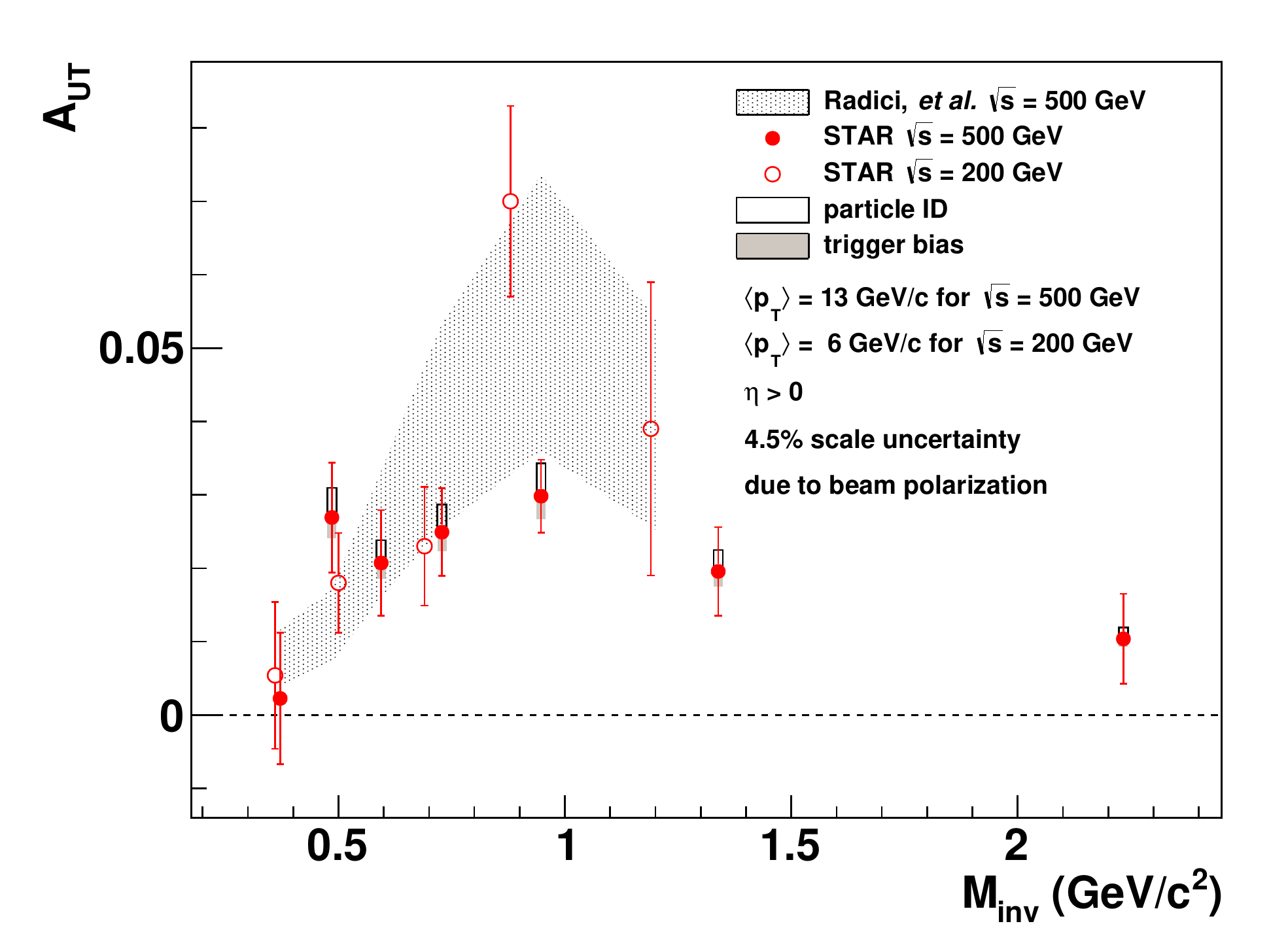}
\vspace*{8pt}
\caption{The azimuthal asymmetry as a function of invariant mass in the highest $p_T$ bin compared with predictions from fits to existing SIDIS and $e^+e^-$ data provided by the same authors as~\cite{Radici:2016lam}. Details on the calculation can be found in~\cite{Radici:2016opu}.
\label{f8}}
%\caption{(top)The azimuthal asymmetry as a function of invariant mass at $\sqrt{s}=500$ GeV and 200 GeV.  Error bars are the statistical uncertainty.  (bottom)The average pair $p_{T}$ in each invariant mass bin, where the $\langle p_{T}\rangle$ were chosen to compare measurements with similar $x_{T}=2p_{T}/\sqrt{s}$.\label{f8}}
\end{figure}    

\section{Analysis}
The azimuthal angles in the scattering system used to calculate the $\pi^{+}\pi^{-}$ azimuthal correlation follow the definition in ref.~\cite{Bacchetta:2004it} and are shown in Fig. \ref{f1}. The scattering plane is defined by the polarized beam direction, $\vec{p}_{beam}$, and the direction of the total momentum of the pion pair, $\vec{p}_{h}$.  The di-hadron plane is defined by the momentum vectors from each pion ($\vec{p}_{h,1}$ and $\vec{p}_{h,2}$) in the pair. The difference vector $\vec{R}=\vec{p}_{h,1}-\vec{p}_{h,2}$ lies in the di-hadron plane. The pions are chosen to be in close proximity to each other in $\eta-\phi$ space with $\sqrt{(\Delta\eta)^{2}+(\Delta\phi)^2} \leq$ 0.7 and the sum of the transverse momenta, $p_{T}$, for each pair is required to be greater than 3.75 GeV/\textit{c}. Throughout the rest of this paper, $p_{T}$ is the transverse momentum of the pion pair and $\vec{p}_{h,1}$ corresponds to the positive pion and $\vec{p}_{h,2}$ to the negative pion. We define the unit vectors $\hat{p}=\vec{p}/|\vec{p}|$. The angle between the scattering plane and the polarization of the incident beam, $\vec{s}_{a}$, is $\phi_{S}$.  The angle between the scattering plane and the di-hadron plane is $\phi_{R}$, which is used to define $\phi_{RS} = \phi_{R}-\phi_{S}$, where $\phi_{R}$ and $\phi_{S}$ are calculated using Eqs. (\ref{cos_s})--(\ref{sin_r}). The angle $\phi_{RS}$ modulates the asymmetry due to the product of transversity and the IFF by sin$(\phi_{RS})$. 

\begin{equation}
cos(\phi_{S}) = \frac{\hat{p}_{beam}\times \vec{p}_{h}}{|\hat{p}_{beam}\times \vec{p}_{h}|}\cdot\frac{\hat{p}_{beam}\times \vec{s}_{a}}{|\hat{p}_{beam}\times \vec{s}_{a}|}
\label{cos_s}
\end{equation}

\begin{equation}
sin(\phi_{S}) = \frac{(\vec{p}_{h}\times \vec{s}_{a})\cdot\hat{p}_{beam}}{|\hat{p}_{beam}\times \vec{p}_{h}||\hat{p}_{beam}\times \vec{s}_{a}|}
\label{sin_s}
\end{equation}

\begin{equation}
cos(\phi_{R}) = \frac{\hat{p}_{h}\times \vec{p}_{beam}}{|\hat{p}_{h}\times \vec{p}_{beam}|}\cdot\frac{\hat{p}_{h}\times \vec{R}}{|\hat{p}_{h}\times \vec{R}|}
\label{cos_r}
\end{equation}

\begin{equation}
sin(\phi_{R}) = \frac{(\vec{p}_{beam}\times \vec{R})\cdot\hat{p}_{h}}{|\hat{p}_{h}\times \vec{p}_{beam}||\hat{p}_{h}\times \vec{R}|}.
\label{sin_r}
\end{equation}

The $\pi^{+}\pi^{-}$ azimuthal correlation observable, $A_{UT}$, is defined in Eq. (\ref{diseqn}), where $P$ is the beam polarization and $N^{\uparrow(\downarrow)}$ is the number of pion pairs when the polarization of the beam is pointing up (down). The combination of different polarization 
directions and detector hemispheres removes luminosity and efficiency dependencies from the asymmetry calculation to leading order~\cite{Ohlsen:1973wf}.
%integrated luminosity when the beam is polarized up divided by the integrated luminosity when the the beam is polarized down.  

$A_{UT}$ is calculated for eight $\phi_{RS}$ bins of equal width in the range $[0,\pi]$, which are then fit with a single-parameter function, $A_{UT}\cdot sin(\phi_{RS})$, to extract the amplitude. The mean reduced $\chi^2$ of all fits is $1.00\pm 0.06$. This procedure is carried out as a function of the pseudorapidity of the pion pair, which is denoted as $\eta$ for the remainder of this report. $\eta>$ 0 is forward with respect to the polarized beam direction.  $A_{UT}$ is also measured as a function of invariant mass, $M_{inv}$, and $p_{T}$.    

%\begin{widetext}
\begin{equation}
A_{UT}\cdot P\cdot sin(\phi_{RS}) =
\frac{\sqrt{N^{\uparrow}(\phi_{RS})N^{\downarrow}(\phi_{RS}+\pi)}-\sqrt{N^{\downarrow}(\phi_{RS})N^{\uparrow}(\phi_{RS}+\pi)}}{\sqrt{N^{\uparrow}(\phi_{RS})N^{\downarrow}(\phi_{RS}+\pi)}+\sqrt{N^{\downarrow}(\phi_{RS})N^{\uparrow}(\phi_{RS}+\pi)}}.
\label{diseqn}
\end{equation}
%\end{widetext}

The scale uncertainty due to the beam polarization in this analysis is 4.5$\%$. We investigated a potential bias of the 
triggered events towards pions that come from quark jets, which could result in an enhancement of the measured asymmetries, 
since gluons are not expected to contribute to transversity.
%Additionally, since quark jets are more collimated than gluon jets, the triggered events are biased towards pions that come from quark jets. This could result in an enhancement of the measured asymmetries, since gluons are not expected to contribute to transversity. 
To investigate this bias, particles produced in $p+p$ simulated events from PYTHIA 6.426 \cite{Sjostrand:2006za} with the Perugia-0 tune \cite{Skands:2010ak}, were processed through a detector simulator (GSTAR package based upon GEANT 3.21/08T \cite{brun}), and then used to estimate the quark/parton ratio of a biased sample over the quark/parton ratio in an unbiased sample.  
In STAR the trigger decision is based on the energy deposit in a defined segment in one of the calorimeters.
We expect therefore that a potential trigger bias effect will be strongest for low $p_T$ parent jets, since at high jet $p_T$ the impact of a shape difference between quark or gluon initiated jets will be negligible for the trigger decision. 
For this reason we investigated the trigger bias as a function of the transverse momentum of the hadron pair. Within our statistical uncertainties, we do not observe a significant trigger bias and thus decided not to correct for this effect. 
Instead, the statistical uncertainty with which one can determine the ratio of the fractions of quark initiated jets in the triggered over the non-triggered sample was assigned as a systematic uncertainty, being $\sim$20$\%$ at low $p_{T}$ and $\sim$5$\%$ at high $p_{T}$. 
%\textcolor{red}{Instead, the precision with which we can determine the ratio of the fractions of quark initiated jets in the triggered over the non-triggered sample is the trigger bias uncertainy, being $\approx$20$\%$ at low $p_{T}$ and $\approx$5$\%$ at high $p_{T}$.} 
%Instead we give the precision with which we can determine the factor $f_\textrm{tb}$ which gives the ratio of the fractions of quark initiated jets in the triggered over the non-triggered sample.
%Tables~\ref{tbl:trig1} and \ref{tbl:trig2} give the $f_\textrm{tb}$ and their uncertainties for each $p_T$ bin. 
Note that the trigger bias does not affect the statistical significance of the measurement because the scaling applies to the asymmetry and its uncertainty equally. Since the trigger efficiency is higher for larger jet energies, the selection of pion pairs might be biased towards lower $z$ pairs. Using the same simulation as for the trigger bias, we estimate this effect to be $\sim$8$\%$ at low $p_{T}$ and $\sim$4$\%$ at high $p_{T}$. 
%Additionally, the kinematics in the triggered sample were compared to the untriggered sample and no significant bias was observed.  

Finally, the pion pair purity previously mentioned was used to estimate the asymmetric asymmetry dilution due to $\pi-K$ and $\pi-p$ pairs and found to be about 15$\%$ and is represented as rectangles above (below) positive (negative) data points in Figs.~\ref{f2}-\ref{f4}. This estimate assumes the $\pi-K$ and $\pi-p$ asymmetries are no larger than the $\pi^{+}-\pi^{-}$ asymmetries and have the same sign. 
%\begin{table}
%\begin{tabular}{c|c}
%$\langle p_T\rangle$ [GeV]& $ f_\textrm{tb} $\\
%\hline
%4.2 & 1.15 $\pm$ 0.87\\
%5.1 &0.83 $\pm$ 0.2 \\
%6.3  & 1.03 $\pm$ 0.21 \\
%8.1  & 1.39 $\pm$ 0.24\\
%10.1  & 1.05 $\pm$ 0.12\\
%12.3  &  1.06 $\pm$ 0.04\\ 
%18.1  &  0.98 $\pm$ 0.04
%\end{tabular}
%\caption{Average $p_T$ and trigger bias scale factors $f_\textrm{tb}$  applicable to the pair $p_T$ binning used in Fig.~\ref{f4}. See main text for details.\label{tbl:trig1}}
%\end{table}
%
%\begin{table}
%\begin{tabular}{c|c}
%$\langle p_T\rangle$ [GeV]& $ f_\textrm{tb} $\\
%\hline
%4.4 &   1.15$\pm $ 0.87 \\ 
%5.2  & 0.83 $\pm$ 0.2\\
%6.3  & 1.03$\pm$ 0.21\\
%8.1  & 1.39 $\pm$ 0.24\\
%13 & 1.04 $\pm$0.06
%\end{tabular}
%\caption{Average $p_T$ and trigger bias scale factors $f_\textrm{tb}$  applicable to the pair $p_T$ binning used in Figs.~\ref{f2},\ref{f3} and~\ref{f5}. See main text for details.\label{tbl:trig2}}	
%\end{table}

\section{Results}
The single spin asymmetry, $A_{UT}$, was measured as a function of $\eta$ for five $p_{T}$ bins. 
It is shown as a function of $\eta$ in Fig. \ref{f2} for the largest $p_{T}$ bin with $\langle p_{T}\rangle$ = 13 GeV/\textit{c}.  
%The data points are centered on the average $\eta^{\pi^{+}\pi^{-}}$ in each bin.  
The other four $p_{T}$ bins have smaller asymmetries compared to the $\langle p_{T}\rangle$ bin in Fig. \ref{f2}.  
Using the particles produced in PYTHIA and processed through GEANT as mentioned previously, the kinematic variables $x$ and $z$ were estimated. 
%, where $z$ is the fractional energy of the parent quark carried by the pion pairs.  
The bottom panel of Fig. \ref{f2} shows $x$ and $z$ as a function of pion pair pseudorapidity. %\textcolor{red}{The $x$ and $z$ values for the other results are provided in the supplementary material.}  
As shown in Fig.~\ref{f2}, a strong rise of the measured signal is observed toward higher $\eta$ 
%in the highest $p_{T}$ bin 
where we reach the highest values of $x$. This is consistent with the expectation that the transversity distribution is largest at high-$x$.

$A_{UT}$ as a function of $M_{inv}$ for $\eta>$ 0 and $\eta<$ 0 is shown in Fig.~\ref{f3} for the five $p_{T}$ bins. For $\eta>$ 0 a significant signal is seen in the highest $p_{T}$ bin, while for $\eta<$0 the values of the asymmetries are significantly smaller as was already shown in Fig.~\ref{f2} for the highest $p_{T}$ bin. For the two highest $p_{T}$ bins and $\eta>$ 0, an enhancement near the $\rho$ mass at mid-$M_{inv}$ is observed. In models of the IFF, this enhancement is expected due to the interference of vector meson decays in a relative $p$-wave with the non-resonant background in a relative $s$-wave~\cite{Radici:2001na}. To test this hypothesis, the same-charge, momentum-ordered ($|\vec{p}_{h,1}| > |\vec{p}_{h,2}|$) asymmetry was calculated and is shown in Fig.~\ref{f5}. This plot shows a significantly smaller asymmetry around the $\rho$ mass compared to the charge-ordered calculation. We note that this suppressed asymmetry can also be explained in single hadron emission models like the Nambu and Jona-Lasinio jet model~\cite{Matevosyan:2013eia} where the parton producing the lower ranked same-charge pion will carry less of the spin information and is more likely to have a transverse momentum direction correlated (instead of anti-correlated) with the higher ranked pion.

$A_{UT}$ as a function of $p_{T}$ for $\eta >$ 0 is shown in Fig. \ref{f4} for five $M_{inv}$ bins.  A significant asymmetry is observed at high $p_{T}$ for $\langle M_{inv}\rangle >$ 0.4 GeV/\textit{c}$^2$.  Though not shown here, the asymmetry as a function of $p_{T}$ for $\eta <$ 0 is small compared to the results for $\eta >$ 0. Supplemental tables containing the numerical results shown in the figures discussed above are available online.

Figure \ref{f8} shows a comparison of a theoretical calculation with the azimuthal asymmetry as a function of the invariant mass measured in $p^{\uparrow}+p$ collisions at $\sqrt{s} = 500$ GeV for the highest $p_{T}$ bin. The gray band represents the range of the 68\% confidence interval of the fit to SIDIS and $e^+e^-$ data~\cite{Radici:2015mwa}. The theoretical prediction for $\sqrt{s} = 500$ GeV has been provided by the authors of reference~\cite{Radici:2016lam}, which was first compared to the STAR results at $\sqrt{s}=200~GeV$ \cite{Adamczyk:2015hri}. 
%These calculations ignore bins in the COMPASS deuteron data ("n.7,8") that drive the $d$-quark transversity to saturate the lower Soffer bound.
The smaller $M_{inv}$ range for the theory band is due to the fact that this specific model calculation has only been performed up to $M_{inv}\approx$ 1.2 GeV/\textit{c}.  The asymmetry comparison shows close agreement within statistical uncertainty between the data and the theory band, which further hints at the universality of the mechanism producing azimuthal correlations in SIDIS, $e^{+}e^{-}$, and $p+p$ data.  These high-precision $\sqrt{s} = 500$ GeV results can further constrain global fits of transversity parton distribution functions to SIDIS, $e^{+}e^{-}$, and $p+p$ data, and in particular, improve the statistical significance for $x >$ 0.1.  

%Fig. \ref{f8} shows a comparison of the azimuthal asymmetry as a function of the invariant mass measured in polarized $p+p$ collisions at both $\sqrt{s} = 500$ GeV and 200 GeV in the top panel, and the average transverse momentum divided by $\sqrt{s}$ for each invariant mass bin in the lower panel.  The 200 GeV result is from STAR data published in \cite{vossen}.  For the 500 GeV data the same invariant mass bin boundaries were used as in Fig. \ref{f3}, however, cuts on the lower $p_{T}$ bound were used in order to compare similar values of $x_{T}=2\langle p_{T}\rangle/\sqrt{s}$ between the 500 GeV and 200 GeV results.  The gray band is the theory band from \cite{tfs}, which is not a fit to the polarized $p+p$ results, but rather an extraction from fitting SIDIS and $e^{+}e^{-}$ data.  The asymmetry comparison shows close agreement within statistical uncertainty between the two center-of-mass energies and the theory band.  These high-precision $\sqrt{s} = 500$ GeV results can further constrain global fits to SIDIS, $e^{+}e^{-}$, and $p+p$ data, and in particular, improve the statistical significance for $x>$0.1$.  

\section{Conclusions}
STAR has measured the first $\pi^{+}\pi^{-}$ transverse spin-dependent azimuthal asymmetries in $p^{\uparrow}+p$ collisions at $\sqrt{s}$ = 500 GeV for several pseudorapidity, invariant mass, and transverse momentum bins.  These data show significant signals at high $p_{T}$ and $M_{inv}$ for $\eta >$ 0.  IFF models predict an enhancement around the $\rho$ mass due to the interference of vector meson decays in a relative $p$-wave with the non-resonant background in a relative $s$-wave.  This prediction is consistent with the data reported in the paper. These data probe transversity at much higher $Q^{2}\approx$400 GeV$^2$ and sample a different mixture of quark flavors compared to the charge weighted coupling in SIDIS.  These results can be used to test the universality of the mechanism producing azimuthal correlations in SIDIS, $e^{+}e^{-}$, and $p+p$.  In the future, a comparison between di-hadron asymmetries with measurements of azimuthal asymmetries of pions in jets will provide further tests of universality and factorization.  Additionally, the high-precision of these results, can further constrain global fits to world data, especially in the region $x >$ 0.1.    

\section{Acknowledgements}
We thank Marco Radici and Alessandro Bacchetta from the Department of Physics at the University of Pavia for helpful discussions and for providing theory curves for this work.
We thank the RHIC Operations Group and RCF at BNL, the NERSC Center at LBNL, and the Open Science Grid consortium for providing resources and support. This work was supported in part by the Office of Nuclear Physics within the U.S. DOE Office of Science, the U.S. National Science Foundation, the Ministry of Education and Science of the Russian Federation, National Natural Science Foundation of China, Chinese Academy of Science, the Ministry of Science and Technology of China and the Chinese Ministry of Education, the National Research Foundation of Korea, GA and MSMT of the Czech Republic, Department of Atomic Energy and Department of Science and Technology of the Government of India; the National Science Centre of Poland, National Research Foundation, the Ministry of Science, Education and Sports of the Republic of Croatia, RosAtom of Russia and German Bundesministerium fur Bildung, Wissenschaft, Forschung and Technologie (BMBF) and the Helmholtz Association.

\bibliography{./bibliography.bib}{}

\end{document}